\documentclass[12pt]{article}

\makeatletter
\newenvironment{indentation}[2]%
{\par \setlength{\leftmargin}{#1} \setlength{\rightmargin}{#2}
  \advance\linewidth -\leftmargin \advance\linewidth -\rightmargin
  \advance\@totalleftmargin\leftmargin \@setpar{{\@@par}}%
  \parshape 1 \@totalleftmargin \linewidth \ignorespaces}{\par}
\makeatother
\usepackage[dvips]{graphics}

\begin{document}

\baselineskip 18pt

\title{\bf A Systematic Algorithm for Quantum Boolean Circuits Construction}
\author{I.M. Tsai\footnote{E-mail : tsai@lion.ee.ntu.edu.tw}
\qquad and \qquad
        S.Y. Kuo\footnote{E-mail : sykuo@cc.ee.ntu.edu.tw}\\
        \\
        \em Department of Electrical Engineering,\\
        \em National Taiwan University,\\
        \em Taipei, Taiwan.\\
       }
\date{\small }
\maketitle

\begin{abstract}
To build a general-purpose quantum computer, it is crucial for
the quantum devices to implement classical boolean logic. A
straightforward realization of quantum boolean logic is to use
auxiliary qubits as intermediate storage. This inefficient
implementation causes a large number of auxiliary qubits to be
used. In this paper, we have derived a systematic way of realizing
any general $m$-to-$n$ bit combinational boolean logic using
elementary quantum gates. Our approach transforms the $m$-to-$n$
bit classical mapping into a $t$-bit unitary quantum operation
with minimum number of auxiliary qubits, then a variation of
Toffoli gate is used as the basic building block to construct the
unitary operation. Finally, each of these building blocks can be
decomposed into one-bit rotation and two-bit control-U gates. The
efficiency of the network is taken into consideration by
formulating it as a constrained set partitioning problem.
\end{abstract}

%---------------------------------------------------------------------------

\section{Introduction}

Since Feynman \cite{Fey82} and Deutsch \cite{Deu85} introduced the
idea and theoretical model of quantum computer in the early
1980's, a great deal of research effort has been focused on the
topic of {\em quantum information science}. The discovery of
Shor's prime factorization \cite{Sho94} and Grover's fast
database search algorithm \cite{Gro96} have made quantum
computing the most rapidly expanding research field recently. For
a quantum algorithm to be useful, it is crucial that the
algorithm should be able to be implemented using quantum gates.
Not long after Deutsch proposed his theoretical model of quantum
computer, he showed that a three-bit quantum gate is universal
and capable of realizing any unitary operation \cite{Deu89}. A few
years later, it was shown \cite{Bar95.1,Div95} that two-bit gates
are sufficient to implement any unitary operation. This makes
experimental implementation of quantum circuits more
practical.\\

Another approach that pushes the computing technology to its
theoretical limits is called {\em nanotechnology}.
Nanotechnology, combining physics and computer science, uses
nanometer scale devices as the fundamental building block of
electronic circuits. Just like a classical computer is built out
of universal classical gates, a quantum computer can be built
using nanoscale quantum gates. Various silicon-based nanoscale
devices have been proposed as candidates for quantum computer
\cite{Los98,Pri98,Kan98,Ima99,Vri99,Ban99}. It is believed that
scalable computation can be achieved using solid state quantum
logic devices. However, to build a general-purpose computer, it
is necessary for these nanoscale quantum devices to be able to
have the ability to implement classical boolean logic.\\

A straightforward realization of quantum boolean logic is to use
auxiliary qubits as intermediate storage. This inefficient
implementation causes a large number of auxiliary qubits to be
used. In this paper, we have derived a systematic way of realizing
any general $m$-to-$n$ bit combinational boolean logic using
elementary quantum gates. Our approach transforms the $m$-to-$n$
bit classical mapping into a $t$-bit unitary quantum operation
with minimum number of auxiliary qubits, then a variation of
Toffoli gate is used as the basic building block to construct the
unitary operation. Finally, each of these building blocks can be
decomposed into one-bit rotation and two-bit control-U gates. The
efficiency of the network is taken into consideration by
formulating it as a constrained set partitioning problem.\\

The rest of this paper is organized as follows. Section 2
describes the relation between permutation and our building block
-- $T(S,R,I)$ gate. The problem and algorithm are defined in
section 3.1 and 3.2, the optimal solution is then derived in
section 3.3 and 3.4. Finally, conclusions are given in
section 4.\\

\section{Gate Representation and Permutation}

A Toffoli \cite{Tof80} gate consists of two {\em control} bits,
$a$ and $b$, which do not change their values, and a {\em target}
bit $c$ which changes its value only if $a=b=1$. The gate can be
written as:
\begin{eqnarray}
{\hat a} & = & a \nonumber \\
{\hat b} & = & b \nonumber \\
{\hat c} & = & (a \wedge b) \oplus c
\end{eqnarray}
where $\oplus$ denotes exclusive-or and $\wedge$ stands for
logical {\bf \small AND}. The three-bit Toffoli gate is a
universal gate. A variation of the three-bit Toffoli gate is
$n$-bit Toffoli gate, indicated by
\begin{equation}
T(S,R,I) \qquad \quad S,R,I \in \{0,1\}^n, \;\;
\Delta(I,\{0\}^n)=1, \;\; S \wedge R = R \wedge I = S \wedge I = 0
\end{equation}
where $\Delta(x,y)$ is the Hamming distance between $x$ and $y$,
$\wedge$ stands for bit-wise logical {\bf \small AND} operation.
The function of a generalized Toffoli gate is similar to that of
a three-bit Toffoli gate. All input bits are left unchanged while
the target bit is inverted conditionally. In the notation shown
above, $S$ and $R$ are indicators that, if expressed in binary
digits, mark the position of control bits. The bits that are set
in $S$ specify the control bits that have to be $1$'s to activate
the logic. Similarly, the bits that are set in $R$ specify the
bits that have to be $0$'s to activate the logic. $I$ simply
represents the target bit to be inverted when the conditions of
$S$ and $R$ are satisfied. Those bits that are not specified in
either $S$, $R$, or $I$ are {\em don't care} bits. Assuming
$n$-bit input $X=x_{n-1}x_{n-2} \cdots x_1x_0$ and target bit
$x_r$, the operation of a $n$-bit Toffoli gate, $T(S,R,I)$, can
be written as:
\begin{eqnarray}
{\hat x_i} & = & x_i, \qquad i=0,1, \ldots ,r-1, r+1, \ldots ,n-1 \nonumber \\
{\hat x_r} & = & (\wedge_{i=0}^{n-1}((s_i \wedge x_i) \vee (r_i
\wedge \bar{x_i}) \vee (\bar{s_i} \wedge \bar{r_i}))) \oplus x_r
\end{eqnarray}
Using this notation, a three-bit Toffoli gate can be represented
as $T_{t}(110,000,001)$, and a control-not gate is written
as $T_{cn}(10,00,01)$.\\

Since the time evolution of any quantum transformation is a
unitary and logically reversible process, thus any quantum boolean
logic can be represented using permutation. A permutation is a
one-to-one and onto mapping from a finite order set onto itself.
A typical permutation $P$ is represented using the symbol
\begin{eqnarray}
P = \left(
\begin{array}{cccccc}
a&b&c&d&e&f\\
d&e&c&a&f&b
\end{array}
\right) \label{permutation}
\end{eqnarray}

This permutation changes $a$$\rightarrow$$d$,
$d$$\rightarrow$$a$, $b$$\rightarrow$$e$, $e$$\rightarrow$$f$,
and $f$$\rightarrow$$b$. The state $c$ stays unchanged. A
permutation can also be expressed as disjoint {\em cycles}. A
cycle includes its members in a list like
\begin{equation}
C=(e_1,e_2, \ldots ,e_{n-1},e_n). \label{cycle}
\end{equation}
The order of the elements describes the permutation. For example,
in Eq.(\ref{cycle}), the cycle takes $e_1$$\rightarrow$$e_2$,
$e_2$$\rightarrow$$e_3$, \ldots ,$e_{n-1}$$\rightarrow$$e_n$, and
finally $e_n$$\rightarrow$$e_1$. The number of elements in a
cycle is called its {\em length}. A cycle with length $1$ is
called a {\em trivial} cycle, which does not change anything. A
cycle of length $2$ is called a {\em transposition}. Using this
notation, the same permutation $P$ shown in
Eq.(\ref{permutation}) can be written as
\begin{eqnarray}
P = (a,d)(c)(b,e,f) = (a,d)(b,e,f)
\end{eqnarray}
Note that a trivial cycle is generally not shown in a permutation.\\

For each permutation $P$, there always exists a permutation
$P^{-1}$ that puts the object back into their place. $P^{-1}$ can
be derived simply by interchanging the two rows of $P$ or, if
cycles are used, reversing the order of the components in each
cycle. A permutation that does not change the order of the
objects is called an {\em identity}, indicated by $E$. If two
permutations, $P_1$ and $P_2$, are performed successively, we
called this the {\em product} of $P_1$ and $P_2$. Following the
convention, we write the first permutation on the right hand side
as $P=P_2P_1$. Clearly, $EP=PE=P$ and $PP^{-1}=P^{-1}P=E$.
Permutations do not commute, i.e. $P_1P_2
\neq P_2P_1$ for general $P_1$ and $P_2$.\\

A quantum boolean logic gate can then be expressed using the
notation described above. For example, a control-not gate is
indicated by $P_{cn}=(10,11)$. Since it changes
$10$$\rightarrow$$11$ and $11$$\rightarrow$$10$, leaving all
other states unchanged.
Similarly, a three-bit Toffoli gate is indicated by $P_t=(110,111)$.\\

\section{Quantum Boolean Logic Construction}

\subsection{Problem Description}

The problem of transforming any $m$-to-$n$ bit combinational
boolean logic into quantum operation can be formalized as follows:\\

{\bf Problem :} Given a classical $m$-to-$n$ bit combinational
boolean logic
\begin{equation}
C: A(\{0,1\}^m) \rightarrow B(\{0,1\}^n),
\end{equation}
and an integer $p$ ($0 \leq p \leq m$), construct a $t$-bit
permutation
\begin{equation}
Q: \Psi(\{0,1\}^t) \rightarrow \Psi(\{0,1\}^t)
\end{equation}
such that for each classical mapping $\alpha=\alpha_0\alpha_1
\cdots \alpha_{m-1} \in A$ ($\alpha_i \in \{0,1\}$) and
$\beta=C(\alpha)=\beta_0\beta_1 \cdots \beta_{n-1}$ ($\beta_i \in
\{0,1\}$), there exist two states $\psi=\psi_0\psi_1 \cdots
\psi_{m-1} \cdots \psi_{t-1} \in \Psi$ and $\phi=\phi_0\phi_1
\cdots \phi_{t-1} \in \Psi$
satisfying:\\
\begin{indentation}{2cm}{1cm}
(1) $\psi_i=\alpha_i$, for $i=0,1, \ldots ,m-1$\\
(2) $\psi_i=0$, for $i=m,m+1, \ldots ,t-1$\\
(3) $Q(\psi)=\phi$\\
(4) $\phi_i=\alpha_i$, for $i=0,1, \ldots ,p-1$\\
(5) $\phi_i=\beta_{i-p}$, for $i=p,p+1, \ldots ,p+n-1$\\
\end{indentation}

The construction process is described in the following sections.\\

\subsection{Building the Quantum Transformation Table}

For any classical combinational boolean logic, a {\em Classical
Transformation Table} can be used to describe the behavior of the
circuits. Taking an $m$-to-$n$ bit circuits as example, a
classical transformation table consists of two parts, a
$2^m$-by-$m$ $\alpha$ table for input, and a $2^m$-by-$n$ $\beta$
table for output. In the $\alpha$ table, there are $2^m$ rows,
numbering from\footnote[4]{We use the notation $A[i][\ast]$ to
denote the $i$-th row, starting from column $0$, all the way to
the end. The notation $A[i][m:n]$, denotes the $i$-th row, from
column $m$ to column $n$. Similar notations are used to denote
column and block.} $\alpha[0][\ast]$ to $\alpha[2^m-1][\ast]$,
and $m$ columns, numbering from $\alpha[\ast][0]$ to
$\alpha[\ast][m-1]$. Similarly, there are $2^m$ rows in the
$\beta$ table, numbering from $\beta[0][\ast]$ to
$\beta[2^m-1][\ast]$, and $n$ columns, numbering from
$\beta[\ast][0]$ to $\beta[\ast][n-1]$. Each row of the $\alpha$
table contains an $m$-bit input pattern, the same row of
the $\beta$ table contains the corresponding $n$-bit output.\\

As in the classical case, a {\em Quantum Transformation Table} is
used to describe a $t$-bit quantum combinational boolean logic. A
quantum transformation table consists of two parts, a
$2^t$-by-$t$ $\psi$ table for input, and a $\phi$ table of the
same size for output. Both tables have $t$ bits in width,
corresponding to $t$ input qubits, numbering from $\psi[\ast][0]$
to $\psi[\ast][t-1]$, and $t$ output qubits, numbering from
$\phi[\ast][0]$ to $\phi[\ast][t-1]$. Similarly, both $\psi$ and
$\phi$ are of length $2^t$, numbering from $\psi[0][\ast]$ to
$\psi[2^t-1][\ast]$, and $\phi[0][\ast]$ to $\phi[2^t-1][\ast]$,
corresponding to all combination of state patterns. Each row of
the $\psi$ table contains a $t$-bit input pattern, the same row of
the $\phi$ table contains the corresponding $t$-bit output.
Because the quantum operation is a reversible unitary
transformation, the $2^t$ rows in the $\phi$
table are simply a permutation of the input patterns.\\

The steps to build the quantum transformation table that based on
the classical circuits is shown below:\\

{\bf Step I. Preserve the input qubits.}

We define the {\em preserved} bits to be the input bits that have
to stay unchanged after the operation, while {\em volatile} bits
are input bits that can be over-written by output bits. Preserved
bits can be used as inputs for other circuits again. Without loss
of generality, assume qubits $0$ to $p-1$ ($0 \leq p \leq m$) are
the bits to be preserved and qubits $p$ to $m-1$ are volatile
bits. Note that $p$ can be zero, in which case no input bit is
preserved. Now prepare two empty tables, $\psi$ and $\phi$, which
are both of size $2^m$-by-$m$. For each row $i$ ( $0 \leq i \leq
2^m-1$), copy $\alpha[i][0:m-1]$ to $\psi[i][0:m-1]$. If $p \neq
0$, also copy the preserved bits from $\alpha[i][0:p-1]$ to
$\phi[i][0:p-1]$, where $0 \leq i \leq
2^m-1$.\\

{\bf Step II. Assign the output qubits.}

Since qubit $0$ to $p-1$ are used to preserve the input bits,
assign qubit $p$ to $p+n-1$ to hold the output bits. Expand the
width of the $\phi$ table whenever needed. For each row $i$ ($0
\leq i \leq 2^m-1$), copy $\beta[i][0:n-1]$ to $\phi[i][p:p+n-1]$. \\

{\bf Step III. Distinguish each output state.}

For a unitary quantum evolution, the quantum transformation table
needs to be one-to-one and onto. For any two patterns $x,y \in
\{0,1\}^{p+n}$ in $\phi$, if $x \neq y$, then set $d=0$, go to
step IV. Otherwise, set
\begin{equation}
d = \lceil \log_{2}M \rceil
\end{equation}
where $M$ is the maximum number of occurrences for a single
pattern. Add extra $d$ columns (numbering from $\phi[\ast][p+n]$
to $\phi[\ast][p+n+d-1]$) to the $\phi$ table. Expand the width of
the $\phi$ table whenever needed. For each row $i$ that has a
repeated pattern, assign a unique $d$-bit pattern to
$\phi[i][p+n:p+n+d-1]$, so that each row in the $\phi$ table has
a different bit pattern. Note that input
bits are good candidates that can be used to distinguish the output patterns. \\

{\bf Step IV. Add auxiliary qubits}

If $m=p+n+d$, no auxiliary qubit is needed. The total number of
qubits, $t$, equals $m$, go to Step V. Otherwise, if $m<p+n+d$,
set $a=(p+n+d)-m$ and add $a$ auxiliary qubits to the $\psi$ table
(numbering from $\psi[\ast][m]$ to $\psi[\ast][m+a-1]$).
Assign these qubits to be all $0$'s. The total number of qubits, $t$, equals $p+n+d$. \\

{\bf Step V. Expand the quantum transformation table}

If auxiliary qubits are used, expand both $\psi$ and $\phi$ tables
to be $2^t$ rows in length. For the $\psi$ table, repeat the
original block $2^a$ times and, for each block, fill in the
auxiliary qubits with a unique $a$-bit pattern. For the $\phi$
table, leave the new entries blank.\\

Based on the constraints derived from the classical boolean
circuit, the quantum transformation table is now partially
constructed. The permutation can be completed simply by filling
in the blanks and make it a one-to-one and onto mapping. However,
to implement the quantum operation efficiently, the permutation
should be carefully selected based on the elementary gate count.
To do this, the gate count evaluation function is introduced in the next section.\\

\subsection{Implementation and Gate Count Evaluation}

The rules that are used to implement an arbitrary permutation is
summarized as follows.\\

{\bf Proposition I.} Given any two states $p$ and $q$ with
$\Delta(p,q)=1$, the transposition $U=(p,q)$ can be implemented
using $T(S,R,I)$, where
\begin{equation}
S=p \wedge q, \qquad R={\bar p} \wedge {\bar q}, \qquad I=p \oplus
q.
\end{equation}

This proposition shows how a transposition of two adjacent states
can be implemented using one $T(S,R,I)$ gate. Note that the
$T(S,R,I)$ gate can be further decomposed into one-bit rotation
and two-bit control-U gates \cite{Bar95.2}.\\

With necessary modification, Proposition I. can be generalized to
implement a transposition of two non-adjacent states as follows:\\

{\bf Proposition II.} Given any two general states $p$ and $q$,
with $\Delta(p,q)=d$, the transposition $U=(p,q)$ can be done
using $2d-1$ adjacent state transpositions.\\

The implementation of a transposition with distance $d$ can be
done in the following way. Assume, in binary expression,
\begin{eqnarray}
p & = & b_0b_1b_2 \cdots b_{t_1} \cdots b_{t_2} \cdots b_{t_{d-1}}
\cdots b_{t_d} \cdots b_{n-1}\\
q & = & b_0b_1b_2 \cdots {\bar b}_{t_1} \cdots {\bar b}_{t_2}
\cdots {\bar b}_{t_{d-1}} \cdots {\bar b}_{t_d} \cdots b_{n-1}
\end{eqnarray}
where $b_i \in \{0,1\}$. Then the transposition $U=(p,q)$ can be
constructed as follows:

(i) Find a list of states, $s_1,s_2, \cdots ,s_{d-1}$, between $p$
and $q$, such that for $1 \leq i \leq d-2$,
\begin{equation}
\Delta(p,s_1)=\Delta(s_i,s_{i+1})=\Delta(s_{d-1},q)=1
\end{equation}
An example of the list is shown as follows:
\begin{eqnarray}
p & = & b_0b_1b_2 \cdots b_{t_1} \cdots b_{t_2} \cdots
b_{t_{d-1}} \cdots b_{t_d} \cdots b_{n-1} \nonumber \\
 s_1 & = &
b_0b_1b_2 \cdots {\bar b}_{t_1} \cdots b_{t_2} \cdots b_{t_{d-1}}
\cdots
b_{t_d} \cdots b_{n-1} \nonumber \\
 s_2 & = & b_0b_1b_2 \cdots
{\bar b}_{t_1} \cdots {\bar b}_{t_2} \cdots b_{t_{d-1}} \cdots
b_{t_d} \cdots
b_{n-1} \nonumber \\
\vdots \nonumber \\
 s_{d-1} & = & b_0b_1b_2
\cdots {\bar b}_{t_1} \cdots {\bar b}_{t_2} \cdots {\bar b}
_{t_{d-1}} \cdots b_{t_d}
\cdots b_{n-1} \nonumber \\
 q   & = & b_0b_1b_2 \cdots {\bar b}_{t_1}
\cdots {\bar b}_{t_2} \cdots {\bar b}_{t_{d-1}} \cdots {\bar b}_{t_d} \cdots b_{n-1} \nonumber \\
\end{eqnarray}

(ii) For the list $p, s_1, s_2, \ldots ,s_{d-1} ,q$, perform the
following adjacent state transpositions:
\begin{equation}
(p,s_1)(s_1,s_2)(s_2,s_3) \cdots
(s_{d-2},s_{d-1})(s_{d-1},q)(s_{d-2},s_{d-1}) \cdots
(s_2,s_3)(s_1,s_2)(p,s_1)
\end{equation}
All the transpositions shown above are performed on two adjacent
states and hence can be implemented using $T(S,R,I)$ gates as
described in Proposition I.\\

Once the transposition of two arbitrary states can be performed. A
general cycle of length $n$ can be constructed. For a trivial
cycle, no gates are needed. For a cycle of length $2$, the
implementation can be easily derived using Proposition I. and
Proposition II. For a cycle of length $n$ ($n \geq 3$), the
following rules are used:\\

{\bf Proposition III.} Given a general cycle $C=(p_0,p_1,p_2,
\ldots ,p_{n-1})$, $C$ can be constructed using $n-1$
transpositions:
\begin{eqnarray}
C & = & (p_0,p_1,p_2, \ldots ,p_{n-1}) \nonumber \\
  & = & (p_1,p_0)(p_2,p_1) \cdots (p_{n-2},p_{n-3})(p_{n-1},p_{n-2})
\end{eqnarray}
Each of these transpositions can be decomposed into
$T(S,R,I)$ gates using Proposition I. and Proposition II.\\

{\bf Proposition IV.} A permutation consists of one or multiple
disjoint cycles. Since disjoint cycles commute, so each cycle
in the permutation can be implemented individually.\\

Given a general cycle $C=(p_0,p_1,p_2, \ldots ,p_{n-1})$, the
distances between any two states $d_i=\Delta(p_i,p_{i+1})$ for
$i=0,1, \ldots ,n-2$, and $d_{n-1}=\Delta(p_{n-1},p_0)$. Assume
$d_m \geq d_i$ for every $i$, the minimum number of $T(S,R,I)$
gates for $C$ can be achieved using the following transpositions:
\begin{equation}
(p_{m+1},p_{m+2})(p_{m+2},p_{m+3}) \cdots
(p_{n-2},p_{n-1})(p_{n-1},p_0)(p_0,p_1)(p_1,p_2) \cdots
(p_{m-1},p_m)
\end{equation}
and the total $T(S,R,I)$ gate count is
\begin{equation}
\Omega_C^T=\sum_{i=0}^{n-1}(2d_i-1)-(2d_m-1)
\end{equation}
Each of these $T(S,R,I)$ gates can be further decomposed into
one-bit rotation and two-bit control-U gates \cite{Bar95.2}. This
results in a network with $\Omega_C^E$ elementary quantum gates.\\

In general, the permutation is a product of disjoint cycles,
\begin{equation}
P=C_0C_1C_2 \cdots C_{n-2}C_{n-1}
\end{equation}
The gate count for $P$ is then
\begin{equation}
\Omega_P^E=\sum_{i=0}^{n-1}\Omega_{C_i}^E
\end{equation}
To build an efficient circuit, the permutation table has to be
constructed with minimum $\Omega_P^E$. This problem is
described in the next section.\\

\subsection{Complete the Permutation with Minimum Gate Count}

Define the digraph $G=(V,E)$, where
\begin{eqnarray}
V & = & \{v_i \mid v_i=\psi[i][\ast], 0 \leq i \leq 2^t-1\}\nonumber\\
E & = & \{(v_s,v_d) \mid v_s=\psi[i][\ast], v_d \cong
\phi[i][\ast], 0 \leq i \leq 2^t-1\}
\end{eqnarray}
The digraph has $2^t$ vertices, corresponding to each of the $2^t$
rows in the $\psi$ table. An edge is defined from $v_s$ to $v_d$
if it is possible for $Q$ to map $v_s$ to $v_d$. The $\cong$ is
used to denote, when only $u$ ($u<t$) bits are specified in
$\phi[i][\ast]$, all states that are {\em compatible} to the
current entry. This results in $2^{t-u}$ edges to be generated
for each of the possible $(v_s,v_d)$ pairs. Filling in the $t-u$
blank bits in the $\phi$ table selects one of the
possible edges and delete others.\\

Using the digraph $G$, the problem is equivalent to finding a set
of disjoint cycles that cover all the vertices in $V$ with minimum
elementary gate count. This is formulated as follows:\\

{\bf Problem :} Given a digraph $G=(V,E)$ and the cost
$\Omega_{C_i}^E$ associated with each cycle $C_i$, find a family
of sets $S=\{S_i \mid S_i=\{v_0^i,v_1^i, \ldots ,v_{n-1}^i\},
v_j^i \in V\}$ and corresponding cycles $C=\{C_i \mid
C_i=(v_0^i,v_1^i, \ldots ,v_{n-1}^i), v_j^i \in V\}$ with minimum
\begin{equation}
\sum_{C_i \in C}\Omega_{C_i}^E
\end{equation}
subject to:
\begin{indentation}{2cm}{1cm}
(1) $\bigcup_{S_i \in S}S_i=V$\\
(2) $\bigcap_{S_i \in S}S_i=\emptyset$\\
\end{indentation}
The problem is essentially a {\em constrained set partitioning
problem}, with each partition being a cycle. There are many set
partitioning problems that have been studied in graph theory
\cite{Ber72} and operations research related works \cite{Hil95}.
A simple but effective algorithm is described here
to demonstrate how the elementary gate count is minimized.\\

{\bf Step I. Enumerate all cycles.}

Given the graph $G(V,E)$ described in the quantum transformation
table, list all cycles $C_i$ ($i=0,1,2,\ldots$) in the graph.
This can be done in the following way:

(i) Select a target edge ($v_\psi,v_\phi$), list all cycles
containing the edge. To find all cycles containing
($v_\psi,v_\phi$), just list all paths from $v_\phi$ to $v_\psi$,
then cycles can be found by concatenating any path from $v_\phi$
to $v_\psi$ with the edge ($v_\psi,v_\phi$).

(ii) Delete the target edge in (i). If there is any edge left in
$G$, go to (i), otherwise all cycles are found. For each cycle
$C_i$, calculate the elementary gate count $\Omega_{C_i}^E$.\\

{\bf Step II. Initialization.}

Let $X=\{x_i\}$ be a $n \times 1$ matrix with $x_i=1$ if $C_i$ is
included in the solution, and $x_i=-1$ if it has been excluded.
Initially set $x_i=0$ for each $i$. Also let $A=\{a_{ij}\}$ be an
$m \times n$ matrix with $a_{ij}=1$ if $v_i
\in C_j$, and $a_{ij}=0$ if $v_i \not\in C_j$.\\

{\bf Step III. Reduction (optional).}

The optional reduction process makes the optimization task easier.
Although there are many effective rules, only three reductions are
described here. Let $S$ be an $n \times 1$ matrix with all
elements set to $1$'s and $R=\{r_i\}=AS$ be the $m \times 1$
matrix that describes the coverage of the vertices.

(i) If $r_i=0$ for any $i$, no solution exists.

(ii) For any $i$, if $r_i=1$ and $a_{ij}=1$, then mark $C_j$ as
included.

(iii) If $C_j$ denotes any cycle that has been included, then all
$C_k$ with $C_k \cap C_j \neq \emptyset$ must be marked as
excluded.

The reduction rules can be applied over again until
no further reduction is possible.\\

{\bf Step IV. Search the optimal solution.}

A depth-first search algorithm is used here to search the optimal
solution.

(i) Set the initial elementary gate count to be $M=\infty$.

(ii) If all vertices are covered, update $M$ and record $X$ in
case $\Omega_{C_i}^E < M$, return. Otherwise, for each $C_j$ that
has not been marked, update $X$ to include $C_j$, apply the
reduction rules as described in Step III, then recursively call
step (ii) with the parameter $X$.

After these steps are done, the selected cycles are recorded
and the optimal gate count is in $M$.\\

\section{Conclusions}

We have derived a systematic way of realizing any general
$m$-to-$n$ bit combinational boolean logic using elementary
quantum gates. Our approach transforms the $m$-to-$n$ bit
classical mapping into a $t$-bit unitary quantum operation with
minimum number of auxiliary qubits. The efficiency of the network
is taken into consideration by formulating it as a constrained
set partitioning problem. This method can be used to transform
classical combinational logic into its quantum version, which is
crucial for a general-purpose quantum computer.

\end{document}